\documentclass[12pt]{article}
\usepackage{latexsym}
\usepackage{amsthm}
\usepackage{graphicx}
\usepackage{epstopdf}
\usepackage{epsfig}
\usepackage{bbm}
\usepackage{amsmath}
\usepackage{amssymb}

\newcommand{\f}[2]{\frac{#1}{#2}}

\def\slashchar#1{\setbox0=\hbox{$#1$}           
   \dimen0=\wd0                                 
   \setbox1=\hbox{/} \dimen1=\wd1               
   \ifdim\dimen0>\dimen1                        
      \rlap{\hbox to \dimen0{\hfil/\hfil}}      
      #1                                        
   \else                                        
      \rlap{\hbox to \dimen1{\hfil$#1$\hfil}}   
      /                                         
   \fi}                                         %
\catcode`@=11
\long\def\@caption#1[#2]#3{\par\addcontentsline{\csname
  ext@#1\endcsname}{#1}{\protect\numberline{\csname
  the#1\endcsname}{\ignorespaces #2}}\begingroup
    \small
    \@parboxrestore
    \@makecaption{\csname fnum@#1\endcsname}{\ignorespaces #3}\par
  \endgroup}
\catcode`@=12

\begin{document}

\baselineskip=18pt

\setcounter{footnote}{0}
\setcounter{figure}{0} \setcounter{table}{0}

\begin{titlepage}
\begin{flushright}
HUTP-06/A0038\qquad \\
hep-ph/0609012\qquad\ \\
\end{flushright}
\vspace{.0in}
\begin{center}

{\Large \bf  
Avoiding an Empty Universe in RS I Models and Large-$N$ Gauge Theories
}

\vspace{0.5cm}

{\bf Jared Kaplan\footnote{kaplan@physics.harvard.edu}, Philip C. Schuster\footnote{schuster@physics.harvard.edu}, and Natalia Toro\footnote{toro@physics.harvard.edu}}

\vspace{.5cm}

{\it Jefferson Laboratory of Physics, Harvard University,\\
Cambridge, Massachusetts 02138, USA}

\end{center}
\vspace{1cm}

\begin{abstract}
\medskip
Many proposed solutions to the hierarchy problem rely on dimensional
transmutation in asymptotically free gauge theories, and these
theories often have dual descriptions in terms of a warped extra
dimension. Gravitational calculations show that the confining phase
transition in Randall-Sundrum models is first-order and
parametrically slower than the rate expected in large-N gauge
theories. This is dangerous because it leads to an empty universe
problem.  We argue that this rate suppression arises from
approximate conformal symmetry. Though this empty universe problem
cannot be solved by using the radion for low-scale inflation, we
argue that if the radion potential is asymptotically free, another
instanton for the RS phase transition can proceed as $e^{-N^2}$. We
also discuss the existence of light magnetic monopoles ($\sim 100$
TeV) as a possible signature of such a phase transition.
\end{abstract}


\end{titlepage}
\tableofcontents \vfill\eject

\section{Introduction}
New confining gauge theories play a central role in many models of
new physics, such as dynamical supersymmetry
breaking, technicolor, and a composite higgs.  If the universe is
heated above the confining temperature, these theories undergo
confining phase transitions as the universe cools.  The vacuum energy
before the phase transition is of order $T_c^4$, so that the universe
in the deconfined phase expands with $H = T_c^2/M_{pl}$.
If the phase transition rate per unit volume is smaller than $H^4$, bubbles
of the new phase never collide, and there is an empty universe problem
\cite{Guth:1982pn}.  From the entropic argument of Section
\ref{sec:confinementGen}, the transition rate per unit volume is $\sim
T_c^4 e^{-N^2}$ for a gauge group of rank $N$.  Thus to avoid the
empty universe problem, we require
\begin{equation}
N \lesssim 2 \sqrt{\log \left( \frac{M_{pl}}{T_c} \right)}.
\label{NBound}
\end{equation}
For $T_c \sim 1$ TeV, $N \lesssim 12$ is required for the phase
transition to complete.

The RS-I model \cite{Randall:1999ee} also undergoes a phase
transition, first discussed by Creminelli et al, in which the TeV
brane nucleates from behind an AdS-Schwarzchild
horizon \cite{Creminelli:2001th,Hawking:1982dh}.  In fact, this is also a confining
phase transition in the CFT dual description, which has rank $N = 4 \pi (M_5
L_{AdS})^{3/2}$\cite{Maldacena:1997re}.  By studying the transition in the weakly coupled
gravitational description, Creminelli et al argue that the transition
is strongly first order and, if the model is weakly coupled, too slow
to complete. If the universe was ever hotter than the weak scale, this
presents a serious problem for weak scale physics that is well
described by an RSI type model.

The analysis of \cite{Creminelli:2001th} employed a generalized
Goldberger-Wise radion stabilization with positive mass squared for
the Goldberger-Wise field.  In the dual picture, this corresponds to
a confinement scale set by competition between a weakly coupled
marginal operator and a slightly irrelevant operator.  The radion
contribution to the tunneling action depends on the mass of the
Goldberger-Wise field and its boundary condition on the TeV brane;
when these parameters are small, the radion action is calculable and
dominates over the gravitational contributions of order $\sim N^2$.

The origin of this scaling behavior is somewhat mysterious in the
gravitational treatment, and it is unclear whether the enhanced
action is peculiar to non-interacting Goldberger-Wise-like
stabilization, or endemic to a more general class of theories.  We
note that because the confining scale in Goldberger-Wise
stabilization is determined by the cancellation of two weakly
coupled operators, the theory has an approximate conformal symmetry
at the confining scale, which is spontaneously broken by
confinement.  The weakly coupled radion is a pseudo-Goldstone boson
of this broken conformal symmetry, and the enhancement of the
brane-nucleation action is nothing more than the usual $S\propto
1/\lambda$ enhancement from the weak radion coupling $\lambda$. This
strongly suggests that the slow transition of
\cite{Creminelli:2001th} is not specific to the Goldberger-Wise
stabilization mechanism, but a result of approximate conformal
symmetry.

This observation suggests a possible instanton for a faster transition
in models, discussed in Section \ref{sec:tunnelNRoll} where the radion
stabilization arises from cancellation of a marginal operator and a
slightly \emph{relevant} one.
Though the theory is approximately conformal at the confining scale
$\mu_{TeV}$, it is far from conformal at the lower scale $\Lambda_S$
at which the slightly relevant operator becomes strongly coupled.
Therefore, if the transition discussed above does not complete, the
hot deconfined theory enters a supercooling phase for $\Lambda_S
\lesssim T \lesssim \mu_{TeV}$.  At $T \sim \Lambda_S$, the radion can
tunnel to $\Lambda_S$ with a rate unsuppressed by weak couplings.  A
short period of inflation follows as the radion rolls classically to
$\mu_{TeV}$, but barring severe fine-tuning, this cannot produce many
$e$-foldings of weak-scale inflation.  Even in this optimistic
scenario, the constraint \eqref{NBound} requires $N \lesssim 10-15$
and hence $L_{AdS} \lesssim 1/M_5$.

In Section \ref{sec:monopoles}, we discuss one additional
phenomenological signature of the Randall-Sundrum picture, namely the
production of TeV-scale magnetic monopoles during the phase
transition. If the Standard Model gauge fields are composite at a
scale $\mu_{TeV}$, then there exist monopoles of mass
$\mu_{TeV}/\alpha$.  In the warped picture, these monopoles are brane
black holes that carry magnetic charge. We give several estimates of
monopole production rates, under various assumptions.  The
phenomenology of such light monopoles is quite different from that of
heavier monopoles, and surprisingly unconstrained.  The Parker bound
derived from galactic magnetic fields is the tightest constraint on
such monopoles.  Most current searches for monopoles in the earth or
monopole flux are insensitive because these monopoles are so light,
but sensitive searches for stopped monopoles could be performed.

\section{The Confinement and Brane-Nucleation Transitions}
We begin by reviewing several well-known properrties of confining phase
transitions in large-$N$ gauge theories.  We argue that (i) the phase
transition is first-order, (ii) the confining temperature $T_c \sim
\Lambda$, and (iii) the transition rate scales as $e^{-S}$, with
action $S \sim N^2$.

The AdS/CFT correspondence relates Randall-Sundrum models with a
weakly stabilized radion to confining gauge theories in which explicit
breaking of conformal symmetry is weak at the confining scale.  These models
serve as a perturbative, calculable check of the large-$N$ scalings reviewed in
\ref{sec:confinementGen} and those derived in Section
\ref{approxConformal}.  In \ref{adscft} we review several properties
of these models, the correspondence, and the brane nucleation phase
transition of RS I discussed in \cite{Creminelli:2001th}.

\subsection{Confining Phase Transitions at Large $N$}
\label{sec:confinementGen} Lattice studies \cite{Lucini:2005vg} show
that for $N\gtrsim 3$ the confining phase transition is first-order,
growing more strongly first order as $N \to \infty$, and that the
critical temperature is of order $\Lambda$.  Indeed, $\Lambda$ is
the only dimensionful scale of the theory.  The approximate
$N$-independence of $T_c/\Lambda$ is less obvious.  In the
deconfined phase, the $N^2$ gluons are independent massless degrees
of freedom, leading to a free energy density $F_{dec} \sim -N^2
T^4$.  The confined phase has $O(1)$ light degrees of freedom and
hence low entropy, but as adjoint bilinears take VEVs of order
$\Lambda$ and scale as $N^2$, the potential energy density is
lowered by $E_{con} \sim -N^2 \Lambda^4$.  The free energies of both
phases are equal at the critical temperature $T_c \sim \Lambda$.

Alternatively, the $N$-independence can be seen by consideration of
confining flux tubes \cite{Polyakov:1978vu}.  A flux tube of length
$r$ and tension $\Lambda$ has energy $\Lambda^2 r$, and there are
$e^{a \Lambda r}$ configurations for such a tube.  The color of the
two charges uniquely determines the $SU(N)$ representation of the flux
tube, so $a$ is $N$-independent.  Thus the partition function for the
flux tubes is
\begin{equation}
Z = \sum_r  e^{\Lambda r (a -\Lambda /T)}
\end{equation}
and we see that $T_c \sim \Lambda$.

Finally, there is an argument that $T_c/\Lambda$ is $N$-independent
\cite{Ishii:2002ys} based on the bag model of
 hadrons.  The transition temperature is the temperature at which
 the thermal density of hadron bags is space-filling, and this occurs
when $n(T) V_{bag}\sim 1$, where $V_{bag} \sim \f{1}{\Lambda^3}$ is
the typical size of hadron bags (with no $N$-dependence) and $n(T)$
is the number density of hadrons.  As long as there is a resonance
with mass $\lesssim \Lambda$, we can estimate the parametric
dependence using $n(T) \sim T^3$, yielding $T_c \sim \Lambda$.

The phase transition rate is controlled by the entropy difference
between the two phases in a region the size of a critical bubble.  The
difference in entropy densities between the two phases at $T_c \sim
\Lambda$ goes as $N^2 \Lambda^3$, and the critical bubble size is
$\Lambda^{-3}$, so that the phase transition rate per unit volume is
$\Gamma \sim \Lambda^4 e^{-c N^2}$ with $c \sim O(1)$.  Our universe,
in the confined phase, is very nearly flat space.  The vacuum energy
difference between the phases, and hence the cosmological constant of
the deconfined phase, is $V_0 \sim N^2 \Lambda^4$, so that for $T
\lesssim \Lambda$ it undergoes inflation with $H \sim
\sqrt{V_0}/M_{pl} \sim N \Lambda^2/M_{pl}$.  If $\Gamma \lesssim
H^4$, bubbles of the confining phase never merge.  To avoid this empty
universe problem, we require
\begin{equation}
N^2 \lesssim \f{4}{c} \ln\left( \frac{M_{pl}}{\Lambda}
\right)
\end{equation}
whenever a gauge group of rank $N$ confines at a scale $\Lambda$
below the inflationary reheating temperature.  Numerically, this requires
$N < 12$ for $\Lambda$ at the weak scale and $c = 1$.

\subsection{AdS/CFT and Confinement}
\label{adscft}
The RS I model provides an explicit AdS/CFT dual description of the
analysis above. The model \cite{Randall:1999ee} consists of a
'slice' of $AdS_5$ bounded by two $3+1$-dimensional branes.  The
bulk metric is given by
\begin{equation}
ds^2 = e^{-2kr} \eta_{\mu \nu} dx^{\mu} dx^{\nu} + dr^2,
\end{equation}
where $k$ is the $AdS_5$ curvature and $M$ is the five dimensional
Planck scale.  The UV brane is located on the hypersurface $r =0$,
while the TeV brane lies at $r = r_c$.  The warped metric means that
the effective UV cutoff for physics on a slice at $r$ is given by $M
e^{-\pi k r}$.  If the Higgs is localized on the TeV brane, the
natural scale for its mass is $M e^{-\pi k r_c}$, while the
effective four-dimensional Planck scale $M_{pl}$ seen by observers
on the TeV brane is given by
\begin{equation}
M_{pl}^2 = \frac{M_5^3}{k} \left(1-e^{-2kr_c} \right)
\end{equation}
By tuning the bulk cosmological constant and brane tensions, $r_c$
(and hence the radion $\mu = k e^{-\pi k r_c}$) can be made a flat direction.
For a realistic model that explains the origin of the hierarchy,
this flat direction must be lifted.  This stabilization can be
achieved by the presence of bulk fields that create an $r_c$
dependent `casimir energy', as in the Goldberger-Wise mechanism
\cite{Goldberger:1999uk}.

The AdS/CFT correspondence clarifies many properties of the RS I
model through a partial dictionary between the conventional
description in bounded AdS space and a four-dimensional, softly
broken conformal field theory coupled to gravity
\cite{Arkani-Hamed:2000ds}. The correspondence is incomplete in that
it is not known precisely which CFT corresponds to the
Randall-Sundrum model with a given set of bulk and brane fields.  It
is useful nonetheless, because so many properties of the gauge
theory are determined by conformal invariance and large-N scaling.
Fields localized to the TeV brane correspond to emergent composite
states of the CFT, and KK modes of bulk fields correspond to
resonances. The bulk fields that stabilize $r_c$ correspond to
slowly running couplings or sources in the CFT.

Above the critical temperature $T_c$ gauge theories deconfine, and
composite states and resonances are no longer the relevant degrees
of freedom.  In the five-dimensional picture, this hot CFT phase
corresponds to an AdS-Schwarzchild geometry\cite{Witten:1998zw}, with metric
\begin{equation}
ds^2 = \frac{\rho^2}{L^2} \f{Z(\rho)}{Z(\rho_{Pl})} d \tau^2 +
\frac{d \rho^2}{Z(\rho)} + \frac{\rho^2}{L^2} \sum_i dx_i^2,
\end{equation}
where $Z(\rho) = 1 - \rho^4/\rho_h^4$, and $\rho_h = \pi L^2 T_h$ is
the coordinate of the black brane.  These coordinates are chosen so
that the induced metric at the Planck brane boundary is identical to
that of pure AdS with the same $L$.  When $T= T_h$, the Hawking
radiation from the black hole horizon is in stable thermal equilibrium
with the bulk.  The confining/deconfining phase transition is between
the AdS-Schwarzchild and RS geometries.  As the hot, deconfined,
expanding universe cools below $T_c$, the confined phase becomes
energetically favorable.

Before turning to more detailed computations, a few comments are in
order.  In the AdS/CFT correspondence, $N^2 = 16 \pi^2 (ML)^3$, so
that large $N$ corresponds to an AdS curvature that is much smaller
than the 5-d Planck scale.  This is a requirement for a sensible
gravitational description.  In the bulk description, the radion $\mu(x) =
k e^{-\pi k r_c(x)}$ is the position of the TeV brane, and
consequently it is a gravitational degree of freedom.  Thus its
kinetic term comes from the Einstein Hilbert action, so it is
accompanied by a factor of $(ML)^3 \propto N^2$.  We will
argue below that in the CFT description the radion is a glueball
state, so that the large-$N$ scaling \eqref{radlagrangian} is consistent
between the two descriptions.

On the CFT side, the black brane can be interpreted as a
space-filling plasma of deconfined CFT matter \cite{Aharony:2005bm}.
A ball of the confined phase corresponds to a bubble of the TeV
brane protruding through the horizon. The wall of the bubble
interpolates between the AdS-S horizon and the TeV brane, but the
two meet at $\mu = 0$ and $T_h = 0$, which is pure AdS space. If the
bubble is big enough, the radion potential takes over and the bubble
expands at the speed of light, corresponding to the radion/glueball
operator condensing out of the vacuum.  As we will discuss below,
the effective field theory breaks down when the temperature on the
TeV brane exceeds the local red-shifted Planck scale, so we only
have control over the regime $\mu > T/(ML)$.

From the perspective of the bulk RS I description it might seem that
the phase transition involves unknown UV physics -- after all, the
transition involves a topology changing GR instanton.  However,
these strong gravitational and stringy effects are merely mocking up
complicated, low energy CFT behavior, and they are not sensitive to
very high energy physics, such as new degrees of freedom at energies
above $T_c$.  Unknown UV behavior does not limit our understanding
of the transition.

Many quantities of interest can be calculated with both the CFT and
RS descriptions.  For the most part we will focus on the CFT side,
so that the level of generality will be obvious. Detailed RS model
estimates were made in
\cite{Creminelli:2001th},\cite{Randall:2006py}; we will consider
them in turn and show how they fit into the general picture of
confining/deconfining transitions.

\begin{figure}[htbp]
\begin{center}
\includegraphics[width=4in]{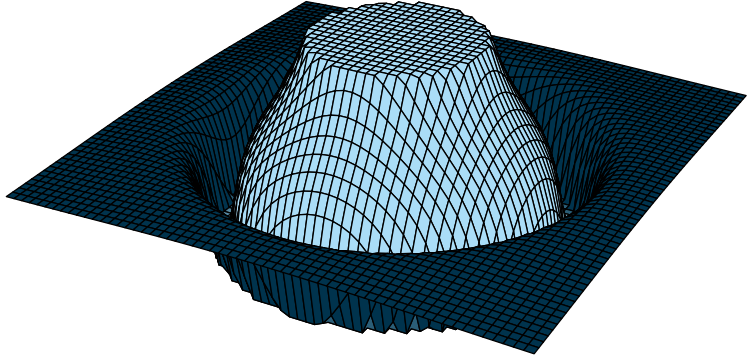}
\caption{An instanton for the TeV-brane nucleation transition in a Randall-Sundrum Model}
\label{instantonfig}
\end{center}
\end{figure}

\section{Confinement with Approximate Conformal Symmetry}
\label{approxConformal} A confining gauge theory is characterized by
its gauge group (which we take to be $SU(N)$ for convenience), its
matter content, a confining scale $\Lambda$, and a strong coupling
scale $\Lambda_{S}$.  In typical asymptotically free theories,
$\Lambda \approx \Lambda_S$. When confinement occurs at the scale of
strong coupling, as in QCD, the approximate conformal symmetry of
the high energy theory is badly broken, so that it plays no role in
the phase transition.  It is possible, however, for confinement to
be induced by a combination of weakly coupled operators perturbing a
CFT.  The confining scale in such a theory, which we call
$\mu_{TeV}$, is parametrically separated from the strong-coupling
scale $\Lambda_S$ ($\Lambda_S\lesssim \mu_{TeV}$)generated by
dimensional transmutation. Though this situation is not generic, it
is of interest both as a calculable regime and because the AdS/CFT
duals of Randall-Sundrum models are of this sort.  At the confining
scale $\mu_{TeV}$, an approximate conformal symmetry governs the
dynamics. In this section we will show that conformal symmetry leads
to  a reduction in the confining phase transition temperature and an
enhancement of the instanton action by powers of the conformal
symmetry breaking spurion.  To do this we must understand the free
energies and relevant degrees of freedom of the two phases.

The approximate conformal symmetry at the scale $\mu_{TeV}$ is
spontaneously broken.  In light of the AdS/CFT duality
between RS-I models and gauge theories with approximate conformal
symmetries, we will refer to the pseudo-goldstone boson \cite{Sundrum:2003yt}
of conformal symmetry breaking as the ``radion'' $\mu$.  Because
the radion is necessarily a singlet under both the $SU(N)$ gauge
symmetry and all other symmetries, it is reasonable to assume that it
is a glueball (or, more generally, an`adjoint-representation-ball'),
and this determines its large $N$ scaling properties. This is borne out in
the specific case of $N=4$ SYM theory, where the radion sets the overall
scale of all moduli \cite{Arkani-Hamed:2000ds}.

In the confined phase there is a radion-glueball condensate at scale
$\mu_{TeV}$, whereas in
the deconfined phase at low temperatures we expect $\langle \mu
\rangle \sim T \ll \mu_{TeV}$\footnote{Since the radion is an operator that creates
a glueball, $\langle \mu \rangle$ can in principle be calculated in the
deconfined phase.  This phase can persist to low temperatures because the
phase transition rate is slow.}.  The radion is an order parameter for the
breaking of the approximate conformal symmetry, and since confinement
can only occur after conformal symmetry is broken, it serves indirectly
as an order parameter for confinement.  For $\mu > \Lambda_S$, the
radion is a good degree of freedom. In this regime, we can calculate
its potential.

By large-$N$ counting, the radion
Lagrangian scales with an overall factor of $N^2$
\cite{Witten:1979kh}, and we can write
\begin{equation} \label{radpotential}
V(\mu) =  N^2 g(\mu) \mu^4,
\end{equation}
\begin{equation} \label{radlagrangian}
L(\mu) = N^2 (\partial \mu)^2 - V(\mu)
.
\end{equation}
As the scale-dependence $\beta$ of the coupling $g(\mu)$
explicitly breaks conformal invariance, we take it to be small at the
minimum of $V(\mu)$, the confining scale $\mu_{TeV}$.  It is important
to note that this implies that $g(\mu_{TeV})$ must be small because
\begin{equation}
V'(\mu_{TeV}) = 0 \ \ \ \implies \ \ \ g(\mu_{TeV}) = -\f{1}{4} \beta \ll 1 ,
\end{equation}
which may be surprising since a $\mu^4$ term with constant coefficient
is conformal.
The entropy of the confined phase is small, so that to a good
approximation
\begin{equation}
F_{conf}(T, \mu) = V_{conf}(\mu);
\end{equation}
 corrections are
considered in Appendix \ref{app:thermal}.

This description is valid only for $\mu \gtrsim \Lambda_S$, for we do
not know what the relevant degrees of freedom are below
the strong coupling scale.  Furthermore, when $T \gg \mu$, many
resonances are thermally excited, and our
description may break down.  For $T \gtrsim N^{2/3} \mu$,
the glueball scattering rate per glueball volume ($\sim \mu^{-3}$)
becomes $O(1)$ and the effective field theory at scale $\mu$ is highly suspect.
The equivalent limit in the dual
5-d gravitational picture, $\mu > T/(ML)$, follows from demanding that
the temperature seen by a TeV brane observer when the brane is at
$\mu$ never exceeds the 5-d Planck scale (the `local Planck scale' is
$M e^{-\pi k r} = \mu (ML)$).  In fact, the local string scale
determined by $M_{s} \sim g_s ^{1/4} M_{pl}$ is an even smaller cutoff
for the effective theory.

\begin{figure}[htbp]
\begin{center}
\includegraphics[width=6.5in]{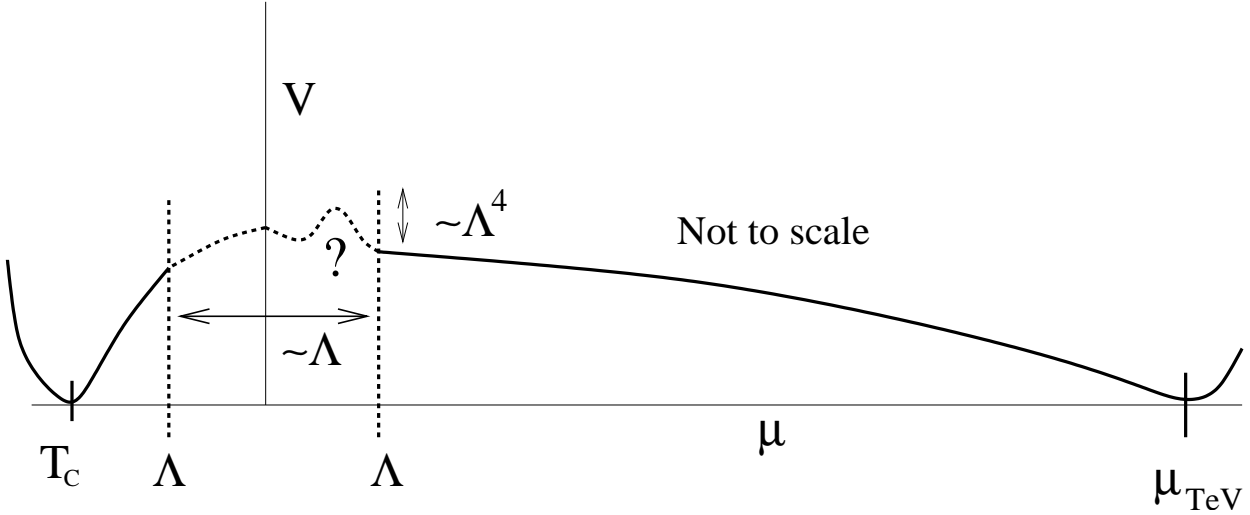}
\caption{Radion potential for the confining phase transition in an approximately conformal field theory.
The region to the left of the Y-axis represents the free energy of the deconfined phase as
a function of $T_h$, while the region to the right represents the free energy of the confined phase
as a function of $\mu$.}
\label{radionfig}
\end{center}
\end{figure}

To study the phase transition, we must calculate the free energy
density as a function of the local plasma temperature $T_h$ in the
deconfined phase.  The free energy density is
minimized when the local temperature $T_h(x)$ is equal to the average
temperature $T$.  Since $T_{\mu \nu}$ is traceless, the hot CFT matter
has the same equation of state as radiation, and we expect $E \propto
N^2 T_h^4$ and $S \propto N^2 T_h^3$.  These two constraints specify
the form of the free energy density up to an overall factor, and this
is in agreement with the answer derived from the AdS-S geometry
in appendix B \cite{Creminelli:2001th},
\begin{equation}
F_{dec}(T_h) = E - T S = \frac{N^2 \pi^2}{8} ( 3 T_h^4 - 4 T T_h^3).
\end{equation}
In equilibrium,  $F_{dec} = -\pi^2 N^2 T^4/8$.  Figure
\ref{radionfig} is a cartoon of the free energy for the system in
both confined and deconfined phases. We have used conformal
invariance in obtaining this expression for $F_{dec}$; conformal
symmetry breaking corrections must be proportional to the spurions
$g(T_h)$ or $g(T)$, which are assumed to be small. These correction
were calculated in \cite{Creminelli:2001th} for a Goldberger-Wise
field in AdS-S, the case relevant to the RS-I model.

The transition temperature and phase transition rate depend on the
relative normalizations and zero-points of $F_{con}$ and $F_{dec}$.  In
an RS-I model, the relative normalization is known because both free
energies are calculated from the same gravitational action, but in the
CFT description they are related by an unknown $O(1)$ factor.  The
zero-point energies are matched at $\mu=T_h=0$, but as
\eqref{radpotential} is strongly coupled for $\mu \lesssim \Lambda_S$,
we can say only that
\begin{equation}
F(T_h=T)-F(\mu_{TeV}) = -\pi^2 N^2 T^4/8 - g(\mu_{TeV}) N^2
\mu_{TeV}^4 + O(N^2 \Lambda_S^4),
\end{equation}
where $g(\mu_{TeV}) < 0$.  The two phases have the same
free energy at the critical temperature
\begin{equation}
T_c \propto [g(\mu_{TeV})]^{1/4} \mu_{TeV}.
\end{equation}
The critical temperature is suppressed by $g^{1/4} \ll 1$ compared
with the strongly coupled result, as claimed.  Below $T_c$, tunneling
between the deconfined and confined phases is allowed.  The action for
the instanton between them involves uncalculable contributions, which
we expect to be $O(N^2)$ as in the conventional confining phase
transition.  But the phase transition also involves tunneling of the
radion from near zero to $\mu_{TeV}$ in the potential
\eqref{radpotential}.  When $\mu$ is rescaled to have a canonical
kinetic term, this has a weak coupling $\lambda = |g(\mu_{TeV})|/N^2$
and the tunneling occurs over a distance $\mu^* = N \mu_{TeV}$ in
field space.  In the thin-walled approximation \cite{Linde:1981zj},
this part of the transition contributes an action
\begin{equation}
S_4 \sim 1/\lambda
\end{equation}
 to the $O(4)$-invariant instanton, and 
\begin{equation}
S_3/T \sim \f{1}{\sqrt \lambda}
\f{\mu^*}{T} \left(1-\left(\f{T}{T_c}\right)^4\right)^{-2}
\end{equation}
to the finite-temperature action for $T \approx
T_c$.  These estimates are accurate to the extent that $\lambda \ll
1$.

The Goldberger-Wise stabilization considered by Creminelli et.\ al.\
has $g(\mu_{TeV}) \approx - \epsilon^{3/2} v_1^2 / N^2$, where $\epsilon
= m_{GW}^2 L_{AdS}^2 /4$ is the anomalous dimension of one of the
Goldberger-Wise operators (taken to be slightly irrelevant), and $v_1$
is the boundary condition for the Goldberger-Wise field on the TeV
brane\cite{Creminelli:2001th}.  The actions they obtain are indeed of
the parametric form expected from our one-parameter description,
underscoring that the effect enhancing the thin-wall action for RS
brane nucleation is precisely approximate conformal invariance at the
scale $\mu_{TeV}$.

\section{A Faster Transition from Strong Coupling in the IR}
\label{sec:tunnelNRoll}
In the gauge theories considered above \ref{approxConformal},
including the Randall-Sundrum model, the confining phase transition is
very slow because the action for tunneling to $\mu = \mu_{TeV}$ is
enhanced by inverse powers of the small conformal symmetry breaking
spurion.  (In the limit of exact conformal symmetry, confinement does
not occur at any finite temperature \cite{Witten:1998zw}).  If this
spurionic operator is marginally irrelevant, then the breaking of
conformal symmetry only gets weaker in the IR and the results of
\cite{Creminelli:2001th} apply. If the deformation is marginally
relevant, conformal invariance becomes strongly broken at an IR scale
$\Lambda_S$.  In Randall-Sundrum models, one usually requires
$\Lambda_S < \mu_{TeV}$, so that explicit breaking of conformal
invariance is weak at the TeV scale.

Near the IR scale $\Lambda_S$, conformal symmetry plays no role in
the dynamics and there is no parametrically light radion.  Thus
tunneling to $\mu \sim \Lambda_S$ is \emph{unsuppressed by small
couplings}, as in a generic asymptotically free theory, though it is
only allowed for $T \lesssim \Lambda_S$.  The essential idea
is that there are two phase transitions -- a confining transition in a
gauge theory, and the spontaneous breakdown of conformal symmetry
in a CFT.  With the instanton of section \ref{approxConformal}, both
transitions occur simultaneously, but in general it should be possible to allow
the universe to cool until the confining transition proceeds in a strongly
coupled regime, after which point the radion rolls to its conformal breaking
minimum without any phase transition at all.  In what follows we consider
whether this process leads to an acceptable spectrum of density perturbations,
and we briefly consider other possibilities for the phase transition.

The confining phase
transition will occur through the nucleation of bubbles of size $\sim
\f{1}{\Lambda_S}$ with a rate per unit volume of $\Gamma \sim
\Lambda_S^4 e^{-cN^2}$, where $c\sim O(1)$.  If the transition is
first order, which we expect for large-N gauge theories \cite{Lucini:2005vg},
then a latent heat of order $\Lambda_S^4$ is released during the
nucleation process. It is energetically favorable for the radion
$\mu$ to relax to $\mu_{TeV}>> \Lambda_S$, which is in
the approximately conformal regime, where
$g(\mu)$ is small and runs slowly.  Though in principle the rolling
radion could lead to many e-foldings of weak-scale inflation, this would
require fine-tuning of the slow-roll parameter $\eta$.  One might
then worry that the late-stage inflation before tunneling destroys
primordial density perturbations.  We will see that this is not a
danger.

It will be useful to introduce a series expansion of the potential
for $\mu$, which we can write quite generally as
\begin{equation}\label{radionPotential}
   V(\mu) = \mu^4 N^2 g(\mu) ,
\end{equation}
where $g(\mu)$ is the slowly running coupling of the conformal
symmetry breaking operator \cite{Rattazzi:2000hs}.  In
the regime where $g$ runs slowly, we can expand $g$ in $V(\mu)$ so
that
\begin{equation}\label{radionPotentialExpanded}
   V(\mu) = \mu^4  N^2  \left(g_0 + g_1 \left(\f{\mu}{\mu_0} \right)^{\epsilon}+... \right).
\end{equation}
where $\mu_0$ is the cutoff, the coefficients $g_i$ depend on
$g(\mu_0)$, and $\f{d \ln g}{d \ln\mu}=\epsilon$. There
is a local extremum at $\mu_{TeV}\approx
\mu_0(\f{-g_0}{g_1})^{\f{1}{\epsilon}}$, exponentially below the
cutoff $\mu_0$, at which the weak scale is stabilized.  This scale
$\mu$ can still be much larger than the IR scale $\Lambda_S$ at
which $g$ gets strong.

It is easy to see that the radion should not be used as a weak scale
inflaton. The flatness of the radion potential over an interval
$\sim \mu_{TeV}$ is protected by the approximate conformal symmetry
above the scale $\Lambda_S$. However, the conformal coupling
$\mu^2R$ between the radion and the Ricci scalar generates a mass
term for $\mu$ of order $H^2\sim \f{V(\mu_{TeV})}{M_{Pl}^2}$, so
that in an inflationary phase $\eta\equiv |\f{V''(\mu)
M_{Pl}^2}{V(\mu)}|\sim \f{H^2m_{Pl}^2}{V(\mu_{TeV})}\sim 1$ and
slow-roll conditions are violated. Tuning of the dynamics near
$\mu\sim \Lambda_S$ is required to make $\eta$ small, and  $60$
$e$-foldings of inflation with nearly scale-invariant density
perturbations after the phase transition seems untenable.

As density perturbations are not generated after the phase
transition, they must not be destroyed during the transition.
Therefore, it is important that the vacuum-energy-dominated
supercooling phase between $T \sim \mu_{TeV}$ and $T \sim \Lambda_S$
does not blow out primordial density perturbations.  In the
supercooling phase below $\mu_{TeV}$, the temperature drops to
$T\sim \Lambda_S$ over $\sim
\ln\left(\f{\mu_{TeV}}{\Lambda_S}\right)$ $e$-foldings of
super-cooling. After nucleation of confining bubbles near the scale
$\Lambda_S$, the confined phase continues to supercool as the radion
degree of freedom $\mu$ relaxes to its weakly coupled minimum.
Because $\eta\sim 1$, we don't expect more than a few $e$-foldings
of inflation after the transition.

It was briefly suggested in \cite{Creminelli:2001th} that deSitter fluctuations
could drive the phase transition.  This means that
we allow the universe to supercool in the deconfined phase until $T \sim H
\sim \mu_{TeV}^2/M_{pl}$, and then hope that the transition
proceeds quickly.  This is not well defined in theories with a strong
coupling scale $\Lambda_S > H$, where we do not know the relevant
degrees of freedom in the deep IR.  In theories
where such a low temperature regime is well-defined, there is no reason
to expect that corrections from deSitter space are qualitatively different
from the thermal corrections considered in appendix B.  Since we have cooled
to $T \sim H$, to avoid the empty universe problem the transition must
proceed very quickly: $T^4 e^{-S} > H^4$ implies $S \sim 0$.
We still have a confining transition in a large N gauge theory, and so
we do not expect such a tiny instanton action.

Before moving on, we should review what has been gained.  We have
shown in this section that it is plausible for an RS model to have a weakly
coupled, approximately conformal description near the TeV scale
without having a confining phase transition that is \emph{parametrically slower}
than the transition in a generic, large N gauge theory.  Unfortunately, since we
need $N < 12$ for gauge theories that become strongly coupled at the
TeV scale, this limits $(ML) < 1$ in RS models, so that even in the most
optimistic scenarios, the extra-dimensional description is barely under control.

\section{Monopoles From Composite Gauge Fields}
\label{sec:monopoles} A number of studies have searched for
monopoles produced in accelerators and stopped in the detectors,
obtaining bounds on the pair production cross-section for monopole
masses up to $800$ GeV (e.g.\ \cite{Abulencia:2005hb}).  If the
compositeness scale $\mu_{TeV}$ is $\sim 1$ TeV or even several TeV,
the monopole mass scale is 100 times greater, well beyond the reach
of these experiments.  Moreover, even if an accelerator were to
reach the monopole mass scale, the production cross-section is
exponentially small.  This suppression follows from the finite size
expected for the monopoles---$1/\mu_{TeV}$, which is $1/\alpha$
times their Compton wavelength.  The two monopoles must be produced
at a spacelike separation $d$ exceeding their size, which implies a
suppression of the production rate by $e^{-m d} \sim e^{1/\alpha}$.
Therefore, the most feasible path to discovering light magnetic
monopoles is not by producing them today, but by detecting
primordial monopoles.

The gauge fields in RS models can be either in the bulk (elementary)
or on the brane(composite).  If the electroweak $SU(2) \times U(1)$
gauge fields are composite at a scale $\mu_{TeV}$, then there is a
magnetic monopole field configuration of mass $\sim
\mu_{TeV}/\alpha$ (this mass scale is simply the magnetic
self-energy for a Dirac monopole solution cut off by new physics at
the length scale $\mu_{TeV}^{-1}$ with magnetic coupling
$1/\alpha$).  In the dual gravitational description, the monopoles
are magnetically charged 4-dimensional black holes at the TeV scale.

\begin{figure}[htbp]
\begin{center}
\includegraphics[width=4in]{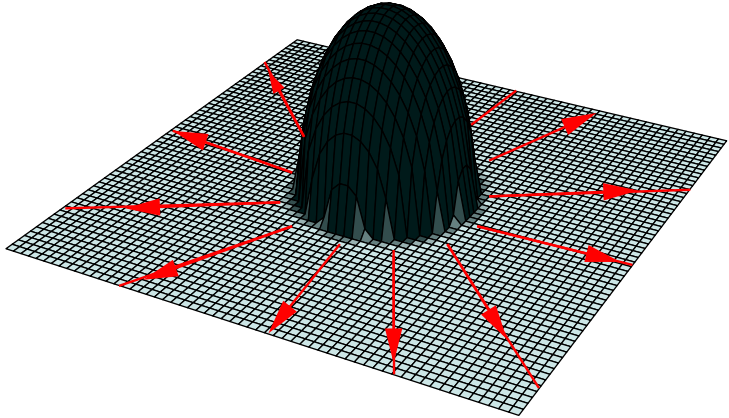}
\caption{A TeV-brane-localized black hole with magnetic charge, with magnetic field lines protruding}
\label{monopolefig}
\end{center}
\end{figure}

One might worry that these monopoles are produced so readily in the
first-order phase transition that they overclose the universe.  This
fate is trivially avoided if the gauge fields are elementary.  But
even when the Standard Model gauge fields are composite, the abundance
of monopoles is low enough that there is no significant constraint.
They may, however, be produced in an abundance close to the Parker
limit.  We present two estimates of their density, and discuss
constraints and possible searches for TeV-scale monopoles.

\subsection{Monopole Formation in the Phase Transition}
A
lower bound on the monopole density (the Kibble density) is obtained
by noting that gauge fields are not coherent between Hubble patches,
so that at least one monopole should exist per Hubble volume at the
transition temperature, and hence
\begin{equation}
(n/s)_{Kibble} \sim \f{H^3}{s} \sim \f{g_* T^6/M_{Pl}^3}{g_{*S} T^3}
\sim 10 \left(\f{T_c}{M_{Pl}}\right)^3 \sim 10^{-47},
\end{equation}
for $T_c \sim 1$ TeV.  We will see that, for a TeV-scale phase
transition, this density is far too low to be observed.

The monopole density may be enhanced by thermal production in the
super-heated regions where bubble walls collide.  The temperature in
these regions is naively $T_{wall} \sim (R/\delta) T_c$, where $R$
is the typical bubble size upon collision and $\delta$ their thickness.  Assuming
the resulting production rate is still less than Hubble, so that the
monopoles are born frozen out, the resulting monopole density is
roughly
\begin{equation}
(n/s)_{superheating} \sim \f{\delta}{R} e^{-2 M/T_{wall}} \sim
\f{\delta}{R} e^{-2 \delta/(R\alpha)}.
\end{equation}
This can be much larger than the Kibble density---indeed, this
mechanism or an alternative must be at work for monopoles to be
visible.

\subsection{Monopole phenomenology}
The thermal velocity dispersion of light monopoles is quite small, so
that their typical velocity relative to galaxies is dominated by the
typical galactic velocity of order $10^{-3} c$.  The Parker bound of
\cite{Parker:1970xv,Turner:1982ag} is applicable, and constrains their
flux.  This bound comes from demanding that the galactic magnetic
field is regenerated by the dynamo effect faster than monopoles
traveling through the galaxy deplete it.  The galactic magnetic field
is $\sim 3 \mu G$.  Light monopoles of charge $h$ that enter the
galaxy at non-relativistic velocities typically gain energy $\Delta
E_1 = hB\ell \sim 10^{11}$ GeV passing through a region of size $\ell
\sim 10^{21}$ cm in which the magnetic field is coherent and of
magnitude $B$.  If a monopole passes through $\sim R/\ell$
uncorrelated regions before escaping the galactic field, where $R \sim
10^{23}$ cm is the radius of the galaxy, then the energy gain of each
monopole is $E_M \sim \sqrt{R/\ell} \Delta E_1 \sim 10^{12}$ GeV.  .
Assuming that the dynamo effect regenerates the galactic magnetic
field on a timescale $\tau \sim 10^8$ years, the flux of magnetic
monopoles must not be sufficient to deplete the field on the same
timescale.  To avoid depleting the galactic magnetic field, then, the
monopole flux must be $\lesssim 10^{-15} cm^{-2} sr^{-1} s^{-1}$, for
$n \lesssim 10^{-23} \mbox{cm}^{-3}$ or $Y \equiv n/s \lesssim
10^{-26}$.

Monopoles reaching the Earth have passed through at least one patch
of coherent galactic magnetic field, from which they gain energy $\sim
10^{11}-10^{12}$ GeV.  The electromagnetic energy loss
rate of monopoles of mass $M=100$ TeV is discussed in
\cite{Wick:2000yc} (see also \cite{Ahlen:1978jy}).  When $E \gg M^2 /m_p \sim 10^{10}$ GeV,
electromagnetic monopole-nucleus scattering transfers an order-1
fraction of the monopole energy to the scattered nucleus, which causes
showering.  Below this energy, the monopole still initiates showering,
but does not lose energy so rapidly, with an energy loss of only $\sim
2 m_p E/M$ per collision.  The resulting atmospheric showers are
visible to cosmic ray detectors, and it has been suggested that
relativistic monopoles could account for super-GZK cosmic rays
\cite{Weiler:1996mu,Wick:2000yc,Anchordoqui:2000mk}, though the
angular and energy profile of observed super-GZK events may be
inconsistent with this proposal \cite{Escobar:1997mr}. At depths of
several kilometers (e.g. the depth of the MACRO detector), monopoles
typically still have $\gamma \sim 10^5$, and they stop at depths of
order $1/10$ the Earth's radius.

The sensitivity of monopole flux limits to 100 TeV monopoles is
unclear.  Because they are very relativistic and have mass $\gg m_e$, the scattering of these
monopoles off electrons is similar to that of ultra-relativistic
cosmic ray muons.  They are not energetic enough to traverse the
Earth, so they contribute only a downward-going flux.  Sensitivity is
then limited by the muon background or the ability to distinguish
monopoles from muons.

 Though the underwater Cerenkov detectors Baikal and Amanda have
measured the flux of downward-going relativistic charged particles,
they have not reported resulting monopole flux bounds
\cite{Aynutdinov:2005sg,Niessen:2001jn,Spiering:2001qd,Orito:1990ny}.  Two analyses
from the MACRO experiment (the combined scintillator-streamer tube
analysis and the track-etch detector) are sensitive to relativistic
monopoles\cite{Ambrosio:2002qq}.  The combined analysis, however, is
only sensitive to $\gamma \lesssim 10$ because more relativistic
monopoles can shower in the detector and may be eliminated by the
selection criteria designed to filter out cosmic-ray
muons\cite{Ambrosio:2001ai}.  The track-etch detector flux limit of
$1.5\,10^{-16} cm^{-2} sr^{-1} s^{-1}$ for monopoles with $\beta \sim
1$ from \cite{Ambrosio:2002qq} may also be insensitive to
ultra-relativistic monopoles because of the increasing mean free path
and because the scattering of such monopoles frequently produces
showers which are less readily distinguished from muons.  Further
analysis of the MACRO data would be necessary to determine its
sensitivity to such light monopoles.

Monopoles that are trapped in the primordial Earth and through-going
monopoles that stop in the vicinity of the crust could both contribute
to a density of monopoles near the surface.  Therefore, limits on the
monopole density per atom of rock (e.g.\ \cite{Jeon:1995rf},
\cite{Ross:1973it} in lunar rock) imply bounds on both flux and
primordial monopole density, though the translation of density limits
is imprecise.  A monopole that crosses through the Earth obliquely and
exits the Earth with kinetic energy below $\lesssim 10$ TeV can lose
this kinetic energy through ionization in traversing the
atmosphere\cite{Wick:2000yc}, after which it is slowed to thermal
velocities, and pulled by the Earth's magnetic field and Brownian
motion it can be bound in magnetized rock\cite{Gamberg:1999tv}.
Only one in $\sim 10^8$ monopoles impinging on the Earth do so at an
angle that leaves $\lesssim 10$ TeV of residual energy, but the
resulting monopole density could still be as high as $\sim 10^{-8}
\mbox{cm}^{-2}$. We note also that the earth's magnetic field $\sim
.3$ G exerts a force $\sim 10^{-3}$ eV/nm on a monopole of unit Dirac
charge, whereas the force of gravity on a monopole of mass $10^5$ GeV
is only $\sim 10^{-11}$ eV/nm.  Thus, it is possible that the stopped
light monopole density is enhanced near the Earth's magnetic poles.
The poles may be interesting places to search for light monopoles.

\section{Conclusions}
Models in which our vacuum emerges from a slow first-order phase
transition can be deadly to cosmology. If the nucleation rate
$\Gamma$ of bubbles of true vacuum is slow compared to $H^4$ at the
transition temperature, then bubbles of true vacuum never collide, and
their interiors are cold and empty.  Such models can only be viable
if the universe never re-heated above the critical temperature, or if
there is second epoch of inflation after the phase transition.

The brane nucleation phase transition in a Randall-Sundrum type-I
model is parametrically slow.  The AdS/CFT correspondence relates this
transition to a confining phase transition in a particular limit of
large-$N$ gauge theory.  This motivates studying the dynamics of the
confining transition in more general large-$N$ theories. The RS-I
instanton is suppressed by two effects. The instanton action $S$ is
proportional to $N^2$, an effect that is endemic to all large-$N$
gauge theories, bounding $N$ in viable weak-scale models to be
$\lesssim 10-14$, with stricter bounds for larger confinement scales.
Furthermore, the calculable part of the instanton of Creminelli,
Nicolis, and Rattazzi in RS-I has a larger action. This enhancement
can be seen to arise from the approximate conformal symmetry at the
scale of confinement.  Indeed, when the radion potential is
asymptotically free, the action for tunneling to the Landau-pole scale
is only $\sim N^2$, as conformal breaking is not a small parameter on
this scale.

Monopoles of mass $\Lambda_S/\alpha$ are always present in theories
in which the electroweak gauge bosons are composite at a scale
$\Lambda_S$.  These monopoles are not produced in the phase
transition at a high enough rate to cause a problem, but they could
be produced at number densities comparable to the Parker limit.
Their phenomenology is quite different from that of GUT-scale
monopoles, and largely uncharted.  If observed, they could offer an
intriguing window on the phase transition.
\section*{Acknowledgements}
Many thanks to Nima Arkani-Hamed, Lisa Randall, Raman Sundrum, Matt
Strassler, and Bobby Acharya for illuminating discussions. P.C.S.
and N.T. are each supported by NDSEG Fellowships. J.K. is supported
by a Hertz Fellowship.
\appendix

\section{The AdS-S Free Energy}

For completeness, we follow \cite{Creminelli:2001th} to derive the free energy
of the AdS-S phase from the gravitational action.

At finite temperature $T$, the AdS-Schwarzchild metric with
a black brane at the Hawking temperature $T_h = T$ solves Einstein's
equations. The black brane has entropy $\propto (ML)^3 \gg 1$, so that at
high temperatures we expect this phase to have lower free energy
than the RS phase.  At very low temperatures, the AdS-S solution
approaches pure Anti-deSitter space, and radion stabilization makes the
RS solution energetically preferred.

We write the AdS-Schwarzchild metric as
\begin{equation}
ds^2 = \frac{\rho^2}{L^2} \f{Z(\rho)}{Z(\rho_{Pl})} d \tau^2 +
\frac{d \rho^2}{Z(\rho)} \frac{\rho^2}{L^2} \sum_i dx_i^2,
\end{equation}
where $Z(\rho) = 1 - \rho^4/\rho_h^4$, and $\rho_h = \pi L^2 T_h$ is the
coordinate of the black brane.  In these coordinates, the induced
metric at the Planck brane boundary is identical to that of pure AdS
with the same $L$.  The thermal compactification of $\tau$ on the
interval $(0,1/T)$ is independent of $\rho_h$.

As in \cite{Creminelli:2001th}, we construct the free energy of the AdS-S
space for all values of $T$, not limited to $T=T_h$. This will have
two pieces---the bulk contribution and a contribution at $T \neq
T_h$ from a conical singularity at the horizon, corresponding to its
lack of thermal equilibrium with the surrounding space.  Though the
Einstein action has a contribution from the Planck brane, it must
vanish because the boundary geometries have been identified.
Physically this makes sense -- UV details should not
be important for the TeV scale free energy.

In the stationary-phase approximation, the free energy can be
approximated as
\begin{eqnarray}
F & = & -T \log \left[ \int Dg_{\mu \nu} \exp \left(-\int_0^{1/T} d \tau
\int d^4x \mathcal{L}_E(g_{\mu \nu})\right) \right] \\
& \approx & T \int_0^{1/T} d\tau \int d^4x \ \mathcal{L}_E(g_{\mu \nu}),
\end{eqnarray}
with Lagrangian density
\begin{equation}
\mathcal{L} = 2 M_5^3 \sqrt{-g} [R + 12 k^2-12 k \delta(\rho
-\rho_{Pl})].
\end{equation}

Since $R=-20 k^2$, the Lagrangian is simply proportional to the
volume of the space, and
\begin{eqnarray}
F_{AdS-S} -F_{AdS} & = & 16 T M^3 k^5 \int_0^{1/T} \left(
\int_{\rho_h}^{\rho_{Pl}} \sqrt{-g_{AdS-S}}  - \int_{0}^{\rho_{Pl}}
\sqrt{-g_{AdS}}\right) \\
 & = & 16 M^3 k^5 \int_{\rho_h}^{\rho_{Pl}} \rho^3 Z(\rho_{Pl})^{-1/2} d\rho -
\int_{0}^{\rho_{Pl}} \rho^3 d \rho\\
 & = & -2 \pi^4 (ML)^3 T_h^4
+ \mathcal{O}\left(\frac{1}{\rho_{Pl}} \right).
\end{eqnarray}

If $T \neq T_h$, we can expand the near-horizon metric in $(\rho -
\rho_h)/\rho_h = y^2/L^2$.  Keeping only the leading terms in y near
the horizon and suppressing the three spatial dimensions, we find a
metric
\begin{equation}
ds^2 = \frac{4 \rho_h^2 y^2}{Z(\rho_{Pl}) L^4} dt^2 + dy^2.
\end{equation}
If $\tau$ is compactified at a radius other than $1/T_h$, this
metric has a conical singularity.  As in [CITE], we regularize this
singularity with a spherical cap of radius $r$, area $2 \pi r^2 (1 -
T_h/T)$, and constant curvature $2/r^2$.  The contribution of this
cap to the free energy is $r$-independent and given by
\begin{equation}
F_{horizon} = 8 \pi^4 (ML)^3 T_h^4 \left( 1 - \frac{T}{T_h} \right).
\end{equation}
Physically, the conical singularity reflects the non-equilibrium
between the black hole and the bulk, which are at different
temperatures, and is proportional to the temperature difference.

Thus, the AdS-S free energy (relative to pure AdS) is given by
\begin{equation}
F_{AdS-S}(T_h) = 6 \pi^4 (ML)^3 T_h^4 - 8 \pi^4 (ML)^3 T T_h^3.
\end{equation}
As expected, the minimum of $F$ is at $T_h = T$.

\section{Thermal Effects and Additional Degrees of Freedom}
\label{app:thermal}
If there are additional elementary degrees of freedom besides the
CFT states, these will contribute to the energy and entropy of both
phases. However, if these fields do not mix significantly with the
CFT, then their properties do not change significantly across the
phase transition, and their effects will be negligable.  An
analogous statement about QCD is that the electron has a very small
effect on the QCD phase transition. In the RS I model, additional
elementary degrees of freedom must reside in the bulk of $AdS_5$ or
on the UV brane (fields on the TeV brane are composite states in the
CFT description).  Bulk and UV brane fields are present as KK modes
in both the AdS-S and RS phases, and they have similar energy and
entropy densities in both phases, so their effects will tend to
cancel.

We assumed above that the number of states in the confining phase
was $O(1)$, which is tiny in comparison to the $O(N^2)$ states in
the deconfined plasma. This is our expectation from the gauge theory
perspective; in general we expect the number of degrees of freedom
to decrease from the UV to the IR \cite{Appelquist:1999hr}, and a parametrically large
number of confined states would be very unusual. However, in RS I
models it is possible to add a large number of confined states by
hand by adding fields to the TeV brane.  From the CFT perspective
this seems rather pathological, but we consider it anyway for
completeness.

Let $g_*$ be the number of confined CFT states, or equivalently, the
number of fields added to the TeV brane.  These states must be light
in order to make a sizable contribution to the entropy.  In the
limit that they are actually massless,
\begin{equation}
\Delta F_{RS} \approx - \frac{\pi^2 g_*}{90} T^4
\end{equation}
Clearly for arbitrarily large $g_*$ these thermal effects could
alter our analysis, but in an RS model we would need $g_* \sim 180
\pi^2 (ML)^3 > 1700$, where the limit comes from the fact that the
5-d Planck scale must be greater than the AdS curvature. It seems
very unlikely that there are this many light composite degrees of
freedom in our universe.

We can also consider the renormalization of the radion potential due
to thermal effects \cite{Weinberg:1974hy}. The radion couples to low-energy
standard model degrees of freedom through \cite{Csaki:2000zn}
\begin{equation}
L_{int} = \frac{\mu^2}{\langle \mu^2 \rangle} T_\nu^\nu
\end{equation}
where $T_\nu^\nu$ is the trace of the energy momentum tensor of the
light degrees of freedom.  This coupling can be derived explicitly
in RS I models, but it also follows directly from the fact that the
radion non-linearly realizes conformal invariance.  For fields such
as massless gauge bosons that are approximately conformal, this
contribution is very small, although it could be significant for
massive fields, for which $\langle T_\nu^\nu \rangle \propto m^2
T^2$ (ignoring terms proportional to small couplings).  For standard
model fields, $m \propto \langle \mu \rangle$ through the higgs
vacuum expectation value, so we find
\begin{equation}
\Delta V(\mu) \sim \sum_{i} y_i \mu^2 T^2 = Y \mu^2 T^2
\end{equation}
where the $y_i$ are effective couplings such as the ratio of the
Higgs mass to $\mu_{TeV}$.  For positive $Y$ this effect will tend
to push the TeV brane away from the Planck brane, making the phase
transition slower.  In the regime where conformal invariance is
softly broken and the transition is under quantitative control, $T_c
< \mu_{TeV}$ and this contribution to the radion potential will be
suppressed compared with the zeroth order $V(\mu)$ coming from
stabilization.

One might still hope that for $Y$ negative, there may be regime
where it is possible to tunnel to some $\mu_T \ll T < T_c$, where
the finite temperature correction dominates over $V(\mu) \propto
\mu^4$.  Unfortunately this is very unlikely -- in order for the
transition to be allowed, we require
\begin{equation}
F_{RS}(\mu_T, T) \approx -Y T^2 \mu_T^2 \leq -2 \pi^4 (ML)^3 T^4 =
F_{AdS-S}(T)
\end{equation}
Since $Y$ must originate as some radion coupling we require $|Y| < 2
\pi^4 (ML)^3$ for consistency of the effective theory, and this
forces $\mu_T > T$. Thus we see that thermal effects cannot drive
the transition or rescue the model.




\begin{thebibliography}{99}

\bibitem{Guth:1982pn}
  A.~H.~Guth and E.~J.~Weinberg,
   ``Could The Universe Have Recovered From A Slow First Order Phase
  Nucl.\ Phys.\ B {\bf 212}, 321 (1983).

\bibitem{Randall:1999ee}
 L.~Randall and R.~Sundrum,
 Phys.\ Rev.\ Lett.\  {\bf 83}, 3370 (1999)
 [arXiv:hep-ph/9905221].

\bibitem{Creminelli:2001th}
 P.~Creminelli, A.~Nicolis and R.~Rattazzi,
 JHEP {\bf 0203}, 051 (2002)
 [arXiv:hep-th/0107141].


\bibitem{Hawking:1982dh}
 S.~W.~Hawking and D.~N.~Page,
 Commun.\ Math.\ Phys.\  {\bf 87}, 577 (1983).


\bibitem{Maldacena:1997re}
 J.~M.~Maldacena,
 Adv.\ Theor.\ Math.\ Phys.\  {\bf 2}, 231 (1998)
 [Int.\ J.\ Theor.\ Phys.\  {\bf 38}, 1113 (1999)]
 [arXiv:hep-th/9711200].


\bibitem{Lucini:2005vg}
  B.~Lucini, M.~Teper and U.~Wenger,
  JHEP {\bf 0502}, 033 (2005)
  [arXiv:hep-lat/0502003].

\bibitem{Polyakov:1978vu}
  A.~M.~Polyakov,
  Phys.\ Lett.\ B {\bf 72}, 477 (1978).

\bibitem{Ishii:2002ys}
  N.~Ishii and H.~Suganuma,
   ``A statistical approach to the QCD phase transition: A mystery in the
  arXiv:hep-ph/0210158.

\bibitem{Goldberger:1999uk}
 W.~D.~Goldberger and M.~B.~Wise,
 Phys.\ Rev.\ Lett.\  {\bf 83}, 4922 (1999)
 [arXiv:hep-ph/9907447].

\bibitem{Arkani-Hamed:2000ds}
 N.~Arkani-Hamed, M.~Porrati and L.~Randall,
 JHEP {\bf 0108}, 017 (2001)
 [arXiv:hep-th/0012148].

\bibitem{Witten:1998zw}
 E.~Witten,
 Adv.\ Theor.\ Math.\ Phys.\  {\bf 2}, 505 (1998)
 [arXiv:hep-th/9803131].

\bibitem{Aharony:2005bm}
  O.~Aharony, S.~Minwalla and T.~Wiseman,
  Class.\ Quant.\ Grav.\  {\bf 23}, 2171 (2006)
  [arXiv:hep-th/0507219].

\bibitem{Randall:2006py}
  L.~Randall and G.~Servant,
  arXiv:hep-ph/0607158.

\bibitem{Sundrum:2003yt}
  R.~Sundrum,
  arXiv:hep-th/0312212.

\bibitem{Witten:1979kh}
  E.~Witten,
  Nucl.\ Phys.\ B {\bf 160}, 57 (1979).

\bibitem{Linde:1981zj}
  A.~D.~Linde,
  Nucl.\ Phys.\ B {\bf 216}, 421 (1983)
  [Erratum-ibid.\ B {\bf 223}, 544 (1983)].

\bibitem{Rattazzi:2000hs}
 R.~Rattazzi and A.~Zaffaroni,
 JHEP {\bf 0104}, 021 (2001)
 [arXiv:hep-th/0012248].


\bibitem{Abulencia:2005hb}
  A.~Abulencia {\it et al.}  [CDF Collaboration],
   ``Direct search for Dirac magnetic monopoles in $p\bar{p}$ collisions at
  Phys.\ Rev.\ Lett.\  {\bf 96}, 201801 (2006)
  [arXiv:hep-ex/0509015].

\bibitem{Parker:1970xv}
  E.~N.~Parker,
  Astrophys.\ J.\  {\bf 160}, 383 (1970).

\bibitem{Turner:1982ag}
  M.~S.~Turner, E.~N.~Parker and T.~J.~Bogdan,
  Phys.\ Rev.\ D {\bf 26}, 1296 (1982).

\bibitem{Wick:2000yc}
 S.~D.~Wick, T.~W.~Kephart, T.~J.~Weiler and P.~L.~Biermann,
 Astropart.\ Phys.\  {\bf 18}, 663 (2003)
 [arXiv:astro-ph/0001233].

\bibitem{Gamberg:1999tv}
  L.~P.~Gamberg, G.~R.~Kalbfleisch and K.~A.~Milton,
  Found.\ Phys.\  {\bf 30}, 543 (2000)
  [arXiv:hep-ph/9906526].

\bibitem{Ahlen:1978jy}
 S.~P.~Ahlen,
 Phys.\ Rev.\ D {\bf 17}, 229 (1978).

\bibitem{Weiler:1996mu}
 T.~J.~Weiler and T.~W.~Kephart,
 Nucl.\ Phys.\ Proc.\ Suppl.\  {\bf 51B}, 218 (1996)
 [arXiv:astro-ph/9605156].

\bibitem{Anchordoqui:2000mk}
  L.~A.~Anchordoqui, T.~P.~McCauley, S.~Reucroft and J.~Swain,
  Phys.\ Rev.\ D {\bf 63}, 027303 (2001)
  [arXiv:hep-ph/0009319].

\bibitem{Escobar:1997mr}
 C.~O.~Escobar and R.~A.~Vazquez,
 Astropart.\ Phys.\  {\bf 10}, 197 (1999)
 [arXiv:astro-ph/9709148].

\bibitem{Aynutdinov:2005sg}
 V.~Aynutdinov {\it et al.}  [Baikal Collaboration],
 arXiv:astro-ph/0507713.

\bibitem{Niessen:2001jn}
 P.~Niessen,
``Search for relativistic magnetic monopoles with the AMANDA
detector,''
http://www.slac.stanford.edu/spires/find/hep/www?irn=6568696 (SPIRES
entry)

\bibitem{Spiering:2001qd}
 C.~Spiering  [AMANDA Collaboration],
 Annalen Phys.\  {\bf 10}, 131 (2001).

\bibitem{Orito:1990ny}
 S.~Orito {\it et al.},
 Phys.\ Rev.\ Lett.\  {\bf 66}, 1951 (1991).

\bibitem{Ambrosio:2002qq}
 M.~Ambrosio {\it et al.}  [MACRO Collaboration],
 Eur.\ Phys.\ J.\ C {\bf 25}, 511 (2002)
 [arXiv:hep-ex/0207020].

\bibitem{Ambrosio:2001ai}
  M.~Ambrosio {\it et al.}  [MACRO Collaboration],
   ``A combined analysis technique for the search for fast magnetic  monopoles
  Astropart.\ Phys.\  {\bf 18}, 27 (2002)
  [arXiv:hep-ex/0110083].

\bibitem{Jeon:1995rf}
  H.~Jeon and M.~J.~Longo,
  Phys.\ Rev.\ Lett.\  {\bf 75}, 1443 (1995)
  [Erratum-ibid.\  {\bf 76}, 159 (1996)]
  [arXiv:hep-ex/9508003].

\bibitem{Ross:1973it}
  R.~R.~Ross, P.~H.~Eberhard, L.~W.~Alvarez and R.~D.~Watt,
   ``Search For Magnetic Monopoles In Lunar Material Using An Electromagnetic
  Phys.\ Rev.\ D {\bf 8}, 698 (1973).


\bibitem{Appelquist:1999hr}
  T.~Appelquist, A.~G.~Cohen and M.~Schmaltz,
  Phys.\ Rev.\ D {\bf 60}, 045003 (1999)
  [arXiv:hep-th/9901109].

\bibitem{Weinberg:1974hy}
  S.~Weinberg,
  Phys.\ Rev.\ D {\bf 9}, 3357 (1974).

\bibitem{Csaki:2000zn}
  C.~Csaki, M.~L.~Graesser and G.~D.~Kribs,
  Phys.\ Rev.\ D {\bf 63}, 065002 (2001)
  [arXiv:hep-th/0008151].

\end{thebibliography}
\end{document}